\documentclass{TAACpaper}
\begin{document}

\selectlanguage{english}

\section{
``Parallel algorithms for problems of cluster analysis with very large amount of data'' 
}

\authors{
Natalya Litvinenko
}

\abstract{
In this paper we solve on GPUs massive problems with large amount of data, which are not appropriate for solution with the SIMD technology. 

For the given problem we consider a three-level parallelization. The multithreading of CPU is used at the top level and graphic processors for massive computing. For solving problems of cluster analysis on GPUs \textbf {the nearest neighbor method} (NNM)  is developed. This algorithm allows us to handle up to \textbf {2 millions} records with number of features up to \textbf {25}. Since sequential and parallel algorithms are fundamentally different, it is difficult to compare the computation times. However, some comparisons are made. The gain in the computing time is about 10 times. We plan to increase this factor up to 50-100 after fine tuning of algorithms.
}

\subsection{Introduction}
Modern NVIDIA video cards have an enormous number of processing cores and CUDA technology. This technology allows us to solve problems, which are impossible to solve with a single-threaded regime \cite{best}. In this work the problem of cluster analysis with large amount of data is solved by the nearest neighbor method. The number of rows achieves 2 millions and the number of features can be up to 25. The main problem of using graphic processors is that they were projected to work with SIMD technology. Unfortunately, only a small part of all tasks can be solved with SIMD technology. In this paper we run several KERNELs, some of them in different threads. This provides us possibility to process in MIMD mode. Each KERNEL is loaded as much as possible in SIMD mode.

\subsection{Analysis of recent results}
Unfortunately, we were not able to find any software for cluster analysis of large amount of data. Such well-known statistic packages as \textbf {SYSTAT, SAS, STATGRAPHICS, SPSS} allow to solve problems of cluster analysis with the amount of data only up to 10000 records.

\subsection{Similar developments}
One of the reasons for developing of this software was information about similar developments at the Novosibirsk State University. This software was applied for solving specific genetics problems. Unfortunately, it was impossible to learn more about this software because this project was closed. Technical performance, potential opportunities of this software are not known for the same reasons.

\subsection{Problem statement}
It is required to solve a massive clustering problem with 2 million records of data by the nearest neighbor method. During constructing the clusters it can be imposed additional conditions (restrictions) that reflect the physical essence of the process. For example, the minimum and maximum numbers of objects in a cluster, the maximum number of clusters, the already built clusters should not join in clusters if the distance between them is greater than the specified one etc. 

During constructing of the algorithm the parts, which can be parallelized by SIMD technology should be processed by separate kernels and number of kernels should depend on maximal loading of GPU. Those parts of the task that can be parallelize with technology MIMD need to process either on the CPU using multithreading or on the GPU using different threads \cite{kepler}.

\subsection{Overall idea of the algorithm}
During solving the given problem, it is assumed that the issues of scaling data, conversion to a common point have already been resolved. By default, the distance between objects during constructing the clusters is defined as follows:
\begin{equation}
\rho_1 =\sqrt{ \sum_j (x_{j,1} - x_{j,2})^2}
\end{equation}

If it is necessary, other metrics could be chosen.

Because of large amount of data, solution of this problem in single-threaded regime, even at fairly good workstation, still will take a very long time - a few days. Therefore, it is necessary to have radically different approach. The first idea is to try to develop a parallel algorithm. This clustering problem could be quite well parallelized. One can immediately see three items that can be parallelized. The first one is the primary processing of data. The second one is construction of distances between all possible pairs of data (by given metric). The third one is selection and sorting of minimum distances between pairs of data.

For solving this problem the GPU and CUDA technologies are used. Graphic processors are mostly focused on the vector processing of data (SIMD). However, it is hard enough to parallelize the algorithms for the SIMD. Additionally, there are quite a lot of computational cores on GPU \cite{sanders2010Cuda}. Therefore, in this paper we attempt to organize a two-level parallelization of the problem. We parallelize the work between some set of cores. Each GPU core solves a single task, i.e. the MIMD technology is used. Inside the kernel (inside of the individual task) there are parts under the SIMD technology \cite{boreskov2012Cuda}.

One of the common problems in parallel data processing on the GPU is the complexity of the maximum and efficient loading of the GPU. There are long threads for parallel processing, but the processing is almost always occurs unevenly on different cores. The output information, as a rule, must be synchronized. Kernels that completed their work first are lying idle.  In practice it is hard to load the GPU even to 50\% and almost impossible to 70-80\%. In this paper we try to shift this problem for special programs (managers), the aim of which is maximal and optimal loading of GPU. 

Several programs, working in parallel in a multithread regime, control the work of GPU cores. These programs are called the second-level managers. In our implementation we use 4 second-level managers. These programs work only with CPU. Their goal is to optimize the work of graphic cores of the processor.

The first level manager controls the work of the second level managers. This first level manager works in a separate CPU thread in parallel with the second level managers. Its main task is to optimize the work of the second level managers. Every second level manager manages  work of several GPU cores.

Quite complicated scheme of data processing described above should solve the problem of optimal loading of GPU cores.

\subsection{Objectives of the main module}
The principal objectives of the main module are: to read program settings from an array of settings; to define auxiliary settings; to check correctness of the data array; to fill the sub-working fields of the array if it is necessary; to run the first level manager; and to perform final computations. 

\subsection{Tasks of the first level manager}
The main tasks of the first level manager are: to ensure the continuous work loading of the second level managers and to process obtained from them results. 

The first level manager has a number of buffers for data, that will be processed by the  second level manager. The number of buffers depends on requirement of ensuring permanent work of GPU cores. In each buffer there is a free portion of data for the first level manager. The buffers are not fixed. The second level manager takes the already prepared portion of data. It should be an excess of prepared buffers because several managers can address to buffers at the same time. The second level manager works parallel and can address to the data at the same time. The practice shows, that for 4 managers it is necessary to have 10-12 buffers. The first level manager should ensure the continuous data availability for work of the second level manager. When the processing of the next portion of data will be finished, every second level manager passes the data for further processing to the first level manager and picks up the next portion of the input data. In order to prevent the situation when the second level manager could not transmit the results, the number of output buffers should be large enough. In our realization there are 12 output buffers.

\subsection{Tasks of the second level manager}
Each manager of the second level has several buffers for the input data. Their number is determined by the number of GPU cores fixed for this manager at the rate of 3 buffers for each core. It is necessary to guarantee permanent work of GPU cores. Buffers with data for cores are not fixed. The core choses another prepared portion of data. Tasks of the second level manager are similar to tasks of the first level manager, but only at a lower level.

\subsection{General description of the algorithm for constructing the set of minimum distances}
It is obviously that to find a pair of points with a minimum distance between them is a difficult and massive operation. Therefore, construction of a cluster on just found pair is unprofitable. It is necessary to process a large number of minimal pairs. However, because of this after processing of the next pair and after unification of two clusters, some of the next pairs will already exist in the joint cluster. Thus the number of united clusters will be significantly less than the number of selected pairs. The number of simultaneously processed pairs is set up by user. Of course, it will be easier to finish the clusterisation process in the first pass, but in this case a lot of minimal pairs will be not used because they are in another cluster and computational resource will be wasted. 

Each GPU core processes the allocated portion of the data and selects P pairs of points, the distance between which is minimal, sorts them by distance, send to the second level manager for further processing, takes the next portion of data. The second level manager selects P minimal pairs among of newly received and already processed pairs. At the end, after processing all the portions of data transferred by the first level manager, P minimal pairs of all data received by the second level manager will be formed. These data will be sent to the first level manager. Four arrays of four controllers are treated in such a way to get P minimal pairs. It will be P minimal pairs around the dataset. Next, a new set of clusters will be constructed. If the process of constructing the clusters is still not over, the process repeats \cite{prokhorenok2010VS}. 

For finding P minimal pairs the data array is logically represented as two blocks of data. The pairs are constructing by selection of an element from each block.

\subsection{Problem settings}
Often during building the clusters the new additional conditions that reflect the physical essence of considered process appear. For example, there are the minimum and maximum limits of objects in a cluster, the maximum number of clusters etc. In this program there are several parameters that allow to include such additional conditions, and to redefine the strategy of building the clusters by the nearest neighbor method: 
\begin{itemize}
 \item KL1 - constructing of clusters will be stopped, if their number is less than KL1;
 \item KL2 - two clusters are not combined into one, if at least one of the considered clusters already has more than KL2 elements. If there was a union of two clusters on the previous step and after this the total number of elements became greater than KL2, the extra elements from the new cluster are not deleted. Therefore, it can be more than KL2 elements in the cluster;
 \item  KL3 - two clusters are not combined into one if the total number of elements in these clusters is greater than KL3. Obviously KL3~$>~$KL2;
 \item KL4 - combine first such group of clusters where at least one has less than KL4 elements.
\end{itemize}

In order to save space and time, only few program settings are described. This is just an illustration of an approach for solving this problem.

\subsection{Development environment}
We used the following hardware and software:

1.	Mainboard - Gigabyte Technology Co., Ltd., Z77MX-D3H, Chipset - Intel;

2.	CPU - Intel(R) Core(TM)  i7-3770 CPU @ 3.40GHz;

3.	Memory - Type: DDR3, Size 16384 Mb;

4.	GPU - NVIDIA GeForce GTX 660 (Kepler architecture, CC - 3.0);

5.	Operating system- Microsoft Windows 7, Ultimate, 32 bit;

6.	Hard disk - 2 Tb;

The following development environment was used:

1.	Programming language - C++;

2.	Development framework - Microsoft Visual Studio 2010;

3.	Work environment for GPU - CUDA 5.5.

\subsection{Analysis of the obtained results}
Development of parallel algorithms for the problem of cluster analysis by the nearest neighbor method in the environment (``CPU" + ``GPU" + ``C++" + ``CUDA") allows to solve difficult problems with large amount of data on cheap hardware.

\subsection{Prospects for further development}
There are different aspects here: 

\textbf {Mathematics.} It is interesting to find theoretical aspects allowing to find easier approaches for solving more complex clustering methods, such as the Ward's method for constructing fast algorithms.

\textbf {Parallel algorithms.} At the first stage we used only a global memory of GPUs. It is interesting to develop algorithms which use faster shared GPU memory. 

\textbf {Programming.} Implementation of the considered clusterisation problems was done in the environment (``CPU" + ``GPU" + ``C++" + ``CUDA"). A part of the task was parallelized by means of CPU (multithreading), and another part by means of the GPU. It is interesting to think about solving this problem in the environment (``OpenMP" + ``GPU" + ``C++" + ``CUDA"). In addition, it is interesting to use cluster systems based on TESLA graphics cards.

\textbf {Applied problems.} The problems of cluster analysis have a great applied significance in solving different problems in biology, genetics, sociology, etc. During solving various problems, often there is a desire to go to other principles of clustering. It is interesting to consider other universally recognized clustering methods that more accurately reflect the considered real processes - Ward's method, the method of full communication (far neighbor) and many others. It is clear that different clustering methods require own approaches for the solution. Development and implementation of the new methods is the first priority in development of this topic. The second priority is the implementation of various limitations that may appear in applied problems.

\subsection{Conclusion}
Conducted investigations have shown that the solving of problems in the environment (``CPU" + ``GPU" + ``C++" + ``CUDA") holds much promise. Parallelization by means of CPU and GPU allow to use SIMD and MIMD technologies. It is greatly expands the range of applied problems. Using of this technology gives the good results. Also, it is not necessary to have expensive hardware and software. So this technology can be widely demanded in various application areas - genetics, biology, sociology, etc.

\bibliographystyle{ieeetr}
\bibliography{NatalyaLitvinenko}

\subsection{Authors}

\author{Natalya Litvinenko}{the 2nd year master, faculty of mechanics and mathematics,
Al-Farabi Kazakh National University, Almaty, Kazakhstan}{n.litvinenko@inbox.ru}

\end{document}